\documentclass{article}
\usepackage{microtype}
\usepackage{graphicx}
\usepackage{booktabs,array,tabularx,longtable,placeins,textcomp}
\usepackage{hyperref}
\usepackage[preprint]{icml2026}
\usepackage{amsmath,amssymb}
\usepackage{xurl}
\usepackage{caption}
\setcitestyle{numbers,square,sort&compress}
\icmltitlerunning{Higher-Scoring Coding Models Did Not Get More Work Done}
\begin{document}
\twocolumn[
  \icmltitle{Higher-Scoring Coding Models Did Not Get More Work Done}
  \begin{icmlauthorlist}
    \icmlauthor{Yining Hua}{harvard}
  \end{icmlauthorlist}
  \icmlaffiliation{harvard}{Harvard University, Boston, MA, USA}
  \icmlcorrespondingauthor{Yining Hua}{yininghua@g.harvard.edu}
  \icmlkeywords{benchmarks, coding agents, natural experiment, pull requests}
  \vskip 0.3in
]
\printAffiliationsAndNotice{}

\begin{abstract}

New coding models advertise themselves on score improvements of coding benchmarks. Yet no study has tested if a model that scores higher on these benchmarks gets more real work done. On June 12, 2026, a U.S. export control directive banned access to Claude Fable 5 for 19 days, which forced users to switch to other models. During this time, most former Fable 5 users switched to Opus 4.8 (11 points lower on SWE-Bench Pro compared to Fable 5). We treat this suspension as a natural experiment, where the people and projects stayed the same and only the model changed, so we could isolate the effect of Fable 5 on user productivity. We tracked GitHub pull requests of 297 Fable 5 users before, during, and after the suspension. We used 8,908 developers who stayed on Opus 4.8 all the time as the control group. We also repeated the comparison across five model upgrades before and after the suspension in 2026 (1,208 developers, 15,442 pull requests). Fitting a difference-in-differences model with developer and date fixed effects, we found that having access to Fable 5 increased the 7-day merge rate of pull requests by only 1.4 percentage points [\textminus{}1.4; +4.3], which was not statistically significant. Similarly, no significant increases of merge rates were found with the upgrades (weighted mean: \textminus{}0.9 percentage points [\textminus{}1.9; +0.2]); all five estimates were negative (95\% intervals included 0, did not differ from each other). We also did not observe a significant increase in the number of pull requests merged per week (+2.4\% [\textminus{}4.9\%; +10.2\%]).

\end{abstract}

\section{Introduction}

When model companies claim their new models have better coding abilities, they almost always show benchmark score improvements on a few coding benchmarks, for example, SWE-Bench Pro~\cite{r6,r20}. The scores are taken as direct evidence for people to quote as better coding abilities, and they are interpreted as the model being more capable of helping developers to get more work done~\cite{r12,r34}.

However, benchmarks are far different than real work. And for coding, where agents have better ability than in any other kind of knowledge work, most benchmark tasks come from issues already solved in public repositories, and they are only checked with existing tests. This is very different than real work, where tasks are often not clearly written, and the code has to fit into the project style~\cite{r33}. And importantly, someone has to be willing to accept it. On GitHub, a code change is proposed as a pull request. Someone would review the pull request of the code change and merge it to add to the project's code base. Maintainers often do not use benchmark tests as a standard for merging pull requests~\cite{r24}. Most benchmarks also do not have evidence on if the scores actually predict how the models perform in real work~\cite{r27,r28,r35}. As a result, no one really knows what an 11-point improvement on a benchmark like SWE-Bench Pro means, and no one knows if it is reflected in real work.

It is hard to measure how much a specific model changes real-work productivity, because the difference in productivity we observe typically does not only come from the model, but also from a lot of other confounders. For example, the people using these models might be different. People who use new models first may be more experienced, so when we compare productivity across developers, we cannot tell if the difference comes from the model or from the people using it. Even if we observe the same people, we do not know if they give the same tasks to different models. For example, some people might use Fable 5 for the hardest tasks and switch to lower-tier models for easier tasks. If they give the harder tasks to the strongest model, the strongest model may look less productive on almost any measure we choose, since harder tasks take longer and are less likely to be merged. Sample composition also matters. In existing studies, most of each agent's pull requests come from a few repositories, which means the merging habits of these repositories themselves might be different, and they decide the number of merges for the whole agent~\cite{r3,r10}. No study that tried to estimate the effect of model upgrades on real work productivity has been able to eliminate these biases.

In this study, we estimate, for the first time, the causal effect of having access to a model that scores higher on benchmarks on developers' productivity, comparing the same developers and varying only their access to the model. In June 2026, a U.S. export control directive banned access to Fable 5 for 19 days~\cite{r5}. During these 19 days, Fable 5 users were forced to switch to lower-tier models, and some of them switched back when access came back. This is a natural experiment where the intervention was set from outside, regardless of users' intentions, allowing us to observe the same people before, during, and after the intervention. We compare Fable 5 users with other developers over the same period using a difference-in-differences design. The main assumption we have is that Fable 5 users' productivity would have changed the same way as other developers' without the suspension.

We collected the data from public GitHub. Claude Code can add a co-author line to the commits of the pull requests it helps create, indicating which model was used~\cite{r19}. We use this line to identify which model was used for each pull request. We measure productivity with merged pull requests. A merged pull request is a patch of code that was finished and added to a larger project, and merge rates are used by existing studies to compare coding agents~\cite{r3,r4}. For each pull request, we check if it was merged within 7 days, and we call the share of pull requests merged within 7 days the merge rate. Then for each developer, we count how many pull requests were merged per week.

Results show that having access to Fable 5 changed the merge rate by only 1.4 percentage points, and the change was not statistically significant (95\% CI \textminus{}1.4 to +4.3). We also analyzed five of the eight model upgrades in 2026 and found that the merge rate did not rise after these upgrades either (\textminus{}0.9 percentage points, 95\% CI \textminus{}1.9 to +0.2). Similarly, developers did not get more pull requests merged per week. These results show that gains on current coding benchmark did not really translate into more finished work.

\section{Related work}

The AIDev dataset collects PRs that coding agents opened on GitHub~\cite{r1}. Later studies discussed how to mine such data~\cite{r2} and why agent PRs are merged or rejected~\cite{r9}. Some studies compared merge rates across agents and task types~\cite{r3,r4}; some other studies counted how many PRs were merged with almost no review~\cite{r10}. These studies do not control for users and models, so the differences between agents they find mix the model, the tool and the people. Other studies were able to fix the tasks by giving people tasks made for the study and comparing their productivity with models of different sizes~\cite{r39}. Nonetheless, since the tasks are assigned, they do not show if developers get more work done in their actual work. Another study compared Python and R projects before and after Copilot's June 2022 model upgrade~\cite{r30}, but it compared projects, not developers. Randomized field experiments~\cite{r11,r21}, a company rollout~\cite{r22} and natural experiments such as Italy's ChatGPT ban~\cite{r29,r40} compare the same developers or randomly assigned ones, but they compare having AI with not having it.

\section{Methods}

\subsection{Study design}

The study answers three questions. First, what is the causal effect of having access to Fable 5 on the merge rate of users' pull requests (PRs)? Second, in model upgrades, was switching to the new model associated with developers' merge rate and the number of PRs merged per week? Third, what does a merge mean?

We answer the first question with a natural experiment through the U.S. export control directive on Fable 5 from June 12, 2026 to July 1, during which no customer had access to Fable 5, only to lower-tier models~\cite{r5}. The strongest of the available models during that period was Opus 4.8, which scored 10.8 points lower than Fable 5 on SWE-Bench Pro~\cite{r6,r20}. We split the PRs of the same Fable 5 users into four phases by creation time: before the release (P0), Fable 5 available (P1), suspended (P2), and after restoration (P3). Developers who never used Fable 5 are used as the control group in a difference-in-differences design (Section 3.5). We assume that without the suspension, the merge rate of Fable 5 users would have changed in the same way as the control group's. We call this the suspension analysis.

We answer the second question with eight model upgrades within the Opus and Sonnet families between February and September 2026 (Table S3). For each developer who switched models, we compare their PRs in the week before and the week after the upgrade. Because developers chose the models themselves, this is only an observational association study. We call this the upgrade analysis.

The third question is descriptive. We extract which account merged the PRs and if anyone else approved them for a sample of 2,019 PRs from the suspension analysis (Section 3.3).

\begin{figure*}[t]\centering\includegraphics[width=\linewidth]{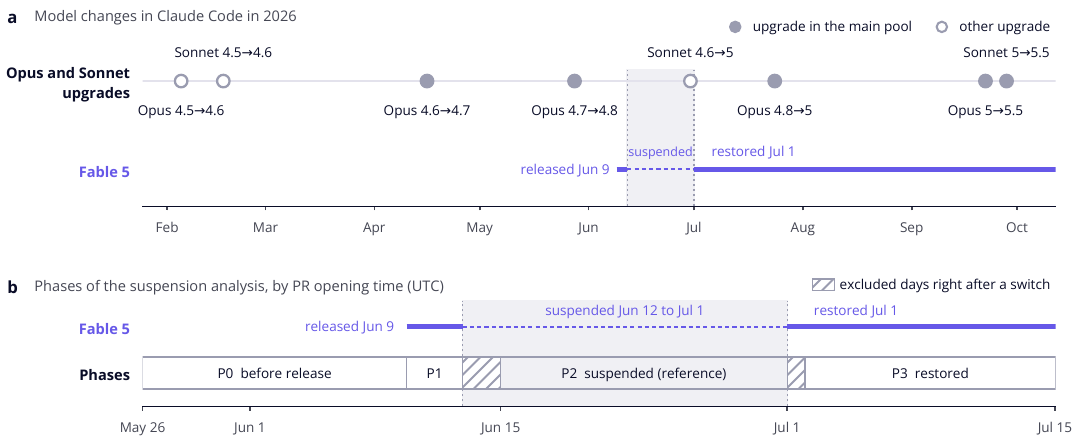}

\caption*{\textbf{Figure 1.} Study design. (a) Model changes in Claude Code in 2026. Dots mark the eight upgrades within the Opus and Sonnet families, and filled dots are the five upgrades in the main analysis. The purple line shows Fable 5, solid when it could be used and dashed when it was suspended. (b) The suspension in a, enlarged, with the four phases of the suspension analysis by PR creation time (UTC). P2 is the reference phase. In both panels, the grey band marks the suspension, from June 12 to July 1. Hatched days, right after Fable 5 was suspended or restored, are excluded.}\end{figure*}

\subsection{Data}

We used the GitHub Search API to collect the public PRs that Claude Code helped create. By default, Claude Code adds the line ``Generated with Claude Code'' to the PR description, and a co-author line with the model name to the commits, for example ``Co-Authored-By: Claude Opus 4.6''~\cite{r19}. When developers include the co-author line in the PR description, e.g., by letting the description reuse the commit message, we can identify which model was used. This method is a common practice to find Claude Code PRs on GitHub~\cite{r1,r37,r38}, and such PRs became common from December 2025 (Table S1). We collected PRs created in the weeks around the Fable 5 suspension and around each model upgrade, between November 20, 2025 and October 5, 2026 (Table S2). We ran the main collection on November 2, 2026 (UTC), by which time the newest PR in the data was four weeks old, so every PR had more than 7 days to be merged.

We kept a PR only if (1) the description names exactly one Claude model, (2) the author is a user account, not a bot, and (3) it was created within the study windows. For the treated group in the suspension analysis, we kept all PRs with a Claude co-author line since this analysis compares phases, not models. To check whether the model's name is correct, we first checked the commits of 200 PRs. In 96\% to 99\% of them, the model in the PR description matched the model in the commits from one day before the PR was created (Appendix B).

For the developers in the suspension analysis, we also collected all the PRs they opened during the study period. For a sample of 2,019 of them, we extracted which account merged them, their reviews and the size of the change (Section 4.3; Appendix D). We release the dataset with the paper (Section 7).

\subsection{Study population}

The treated group in the suspension analysis contains 297 developers who opened at least one PR that named Fable 5 during the three days Fable 5 was available (P1). The control group contains 8,908 developers who opened PRs naming Opus 4.8 between May 26 and July 15, 2026, but never opened anything naming Fable 5.

The upgrade analysis uses developers who switched models at one of the five upgrades in the main analysis, which means they opened PRs with the old model in the week before the upgrade and with the new model in the week after (1,208 developers, 15,442 PRs; Appendix A). A developer who switched at several upgrades is counted once for each upgrade. We also use 944 developers who used another Claude model in both weeks and did not switch (8,569 PRs), only for weekly merged PRs and for the calendar comparison (Section 3.5).

The third question uses 2,019 PRs from 194 developers in the suspension analysis, including all 1,019 PRs by outside authors in the treated group and 1,000 randomly sampled PRs by inside authors from all four phases.

\subsection{Outcomes}

We use two merge-based main outcomes: (1) 7-day merge rate, defined as the share of PRs merged within 7 days of creation; (2) weekly merged PRs, defined as the number of PRs a developer opened in a week that were merged within 7 days. Field studies of AI coding assistants also use the number of completed or merged PRs as the main productivity measure~\cite{r21,r22}. We discuss what merges can and cannot capture in Appendix A. We estimate weekly merged PRs only in the upgrade analysis, because the suspension analysis has no reliable control group for this outcome (Appendix F). There are four secondary outcomes: (1) merged within one hour of creation; (2) closed without being merged; (3) number of comments on the PR; and (4) reverted within 30 days after merge, i.e., a later PR in the same repository with ``revert'' in its title refers to it by its title or number (Appendix F). We also look at automated checks, i.e., the tests, builds and other checks that GitHub runs on a PR, and define a check failure as a PR with at least one check that failed or timed out (Appendix A).

\subsection{Statistical analysis}

For the suspension analysis, we fit a difference-in-differences model at the PR level:

\[Y_{it}=\alpha_i+\gamma_t+\beta_0\,P0_{it}+\beta_1\,P1_{it}+\beta_3\,P3_{it}+\varepsilon_{it}\]

$Y_{it}$ denotes a PR that developer i opened on day t. It is 1 if it was merged within 7 days, and 0 otherwise. $\alpha_i$ is the developer fixed effect. It adjusts for differences between developers, as long as these differences stay the same over time. $\gamma_t$ is the calendar-day fixed effect, which adjusts for changes shared by all developers on the same day. P0, P1 and P3 are indicators of the treated group's PRs in each phase, with P2 (suspension) as the reference. The main estimate is the reversal contrast $(\beta_1+\beta_3)/2$, i.e., the average difference in merge rate between the two phases with Fable 5 and the suspension. $\beta_0$ is a placebo. Standard errors are clustered by developer.

For the upgrade analysis, we fit $Y_{ip}=\alpha_i+\delta\cdot \mathit{New}_{ip}+\varepsilon_{ip}$ for each upgrade, where $\mathit{New}_{ip}$ is 1 if PR p was opened with the new model in the week after the upgrade. $\delta$ is the change in merge rate of the same developers, and the five upgrades are pooled with inverse-variance weighting. We also replace $\alpha_i$ with developer $\times$ repository fixed effects. For weekly merged PRs, pooled over six upgrades (Appendix A), we compare the change in log(1 + merged PRs) between developers who switched models and developers who did not switch. Sensitivity analyses are in Table S5.

We report estimates with 95\% confidence intervals (crossing 0 means statistical insignificance). We do not adjust for multiple comparisons, and for the sensitivity analyses we look at the size and direction of the estimates across settings instead of statistical significance, to avoid p-hacking~\cite{r41}.

\section{Results}

Before Fable 5 release, distributions of the treated and control groups characteristics were highly similar (all standardized differences between \textminus{}0.18 and +0.15; Table 1). In the upgrade analysis, developers who switched models opened more PRs in other people's repositories than those who did not (outside authors, 23.5\% vs 11.7\%; Table 1). After model switching, the same developers also opened more PRs in their own repositories (from 64.3\% to 74.5\%; Table S6).

We observe that Fable 5 users stopped using Fable 5 within a day of the directive. During the three days Fable 5 was available (P1), 82.3\% of their PRs named Fable 5. During the suspension (P2), only 0.9\% did, and 74.6\% named Opus 4.8. After access came back (P3, July 2 to 15), only 26.9\% of their PRs named Fable 5 again. From P1 to P3, the share of PRs opened in the authors' repositories stayed between 69\% and 72\%, and the type of work did not change meaningfully (Appendix F).

\begin{table*}[t]\centering

\caption*{\textbf{Table 1.} Characteristics of the study population. The left columns show the suspension analysis, using all PRs that treated developers and sampled control developers opened before the Fable 5 release (P0, May 26 to June 9, 2026), whether or not they name a model. The right columns show the upgrade analysis, using PRs opened in the week before each of the five upgrades in the main analysis by developers who switched models and by developers who used another Claude model in both weeks and did not switch. A developer who appears at more than one upgrade is counted once for each upgrade. The standardized difference is defined as the difference in means divided by the pooled standard deviation, and PRs per developer and description length are on the log(1 + x) scale. Outside authors neither own the repository nor are members or collaborators. The full table, with more characteristics, the weighted control group, the treated group across the four phases, and the developers who switched before and after the upgrades, is in Table S6. Section 3.5 explains how we handle the differences between the groups.}

\small

\begin{tabularx}{\textwidth}{>{\raggedright\arraybackslash}p{4.4cm} *{6}{>{\centering\arraybackslash}X}}\toprule

 & \multicolumn{3}{c}{Suspension (P0)} & \multicolumn{3}{c}{Upgrades (week before)}\\

 & Treated & Control & Std. diff. & Switched & Did not switch & Std. diff.\\

\midrule

Developers with PRs & 244 & 315 &  & 1,208 & 944 & \\

PRs & 21,377 & 30,630 &  & 7,829 & 4,502 & \\

PRs per developer, median (IQR) & 34 (10--88) & 35 (10--89) & \textminus{}0.06 & 2 (1--6) & 2 (1--5) & +0.12\\

Merged within 7 days (\%) & 92.7 & 91.5 & +0.05 & 95.0 & 94.2 & +0.03\\

Merged within one hour (\%) & 73.9 & 71.5 & +0.05 & 77.8 & 77.4 & +0.01\\

Closed without merge (\%) & 6.0 & 6.8 & \textminus{}0.03 & 3.9 & 4.1 & \textminus{}0.01\\

Opened in the author's own repository (\%) & 65.9 & 67.3 & \textminus{}0.03 & 64.3 & 75.5 & \textminus{}0.25\\

Outside author (\%) & 18.8 & 16.8 & +0.05 & 23.5 & 11.7 & +0.31\\

Title starts with ``feat'' (\%) & 23.5 & 31.5 & \textminus{}0.18 & 18.3 & 12.4 & +0.17\\

Description length, median (characters) & 1,393 & 1,259 & +0.15 & 1,267 & 1,108 & +0.11\\

\bottomrule\end{tabularx}

\end{table*}

\subsection{Losing and regaining Fable 5 barely changed the merge rate}

\begin{figure*}[t]\centering\includegraphics[width=\linewidth]{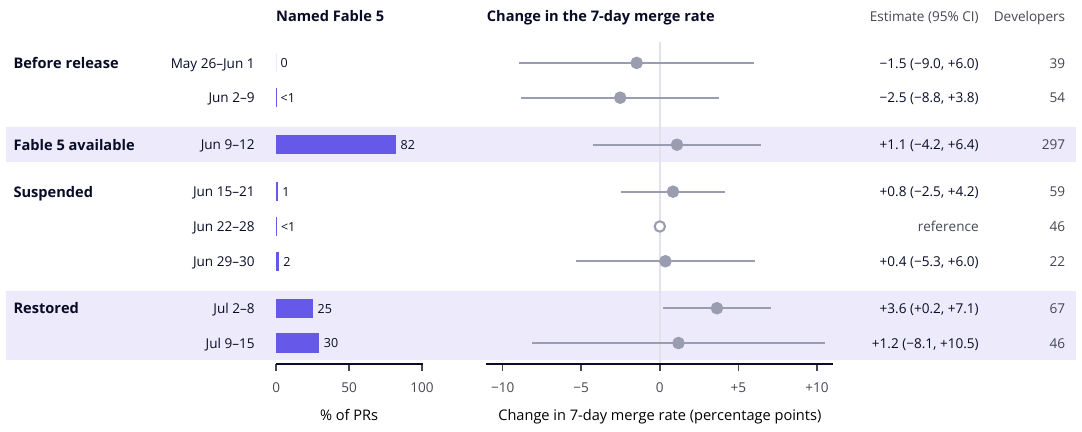}

\caption*{\textbf{Figure 2.} Use of Fable 5 swung from 82\% to under 1\% and back to about 30\%, while the same developers' merge rate stayed flat. Left, share of the treated group's PRs that named Fable 5 in each period. Right, treated $\times$ period coefficients for the 7-day merge rate from the linear probability model with developer and calendar-day fixed effects, with 95\% intervals clustered by developer. The reference period is June 22 to 28, inside the suspension. The last column gives the treated developers active in each period. Shaded rows had Fable 5 available, and the excluded days (June 12, 21:00 to June 15, and July 1) are not shown. Claude Code PRs, May 26 to July 15, 2026.}\end{figure*}

Figure 2 shows that when Fable 5 was available (P1 and P3), the merge rate among the treated group was only 1.4 percentage points (95\% CI \textminus{}1.4 to +4.3; Table 2) higher than during the suspension, i.e., having access to Fable 5 increased the merge rate by at most 4.3 percentage points. The placebo, which compares the weeks before the release (P0) with the suspension, was \textminus{}2.6 percentage points (\textminus{}6.2 to +1.1).

We got similar results in other settings (Table S5). With other weightings, a weekly linear trend, a refit on all PRs, or other outcome definitions, the change was between \textminus{}0.1 and +3.7 percentage points. Only one setting gave a statistically significant result. With all PRs and equal weight per developer, the change was +3.6 percentage points (+0.1 to +7.1). When we gave each developer equal weight, the estimate was larger but much less precise (Appendix A). Developers did not use Fable 5 for all of their PRs when it was available. For a PR that used Fable 5, we estimated the change at +2.7 percentage points (\textminus{}2.6 to +8.0), which was not statistically significant. Comparing the PRs that named Fable 5 with the same developers' other PRs gave the same result (+2.7, \textminus{}0.7 to +6.1; Appendix D).

\subsection{Switching to a new model did not increase the merge rate or weekly merged PRs}

Pooled over the five upgrades, the merge rate of developers who switched changed by \textminus{}0.9 percentage points (95\% CI \textminus{}1.9 to +0.2; Figure 3b and Table 2), which was also not statistically significant, i.e., switching to a new model increased the merge rate by at most 0.2 percentage points. Comparing PRs only within the same developer and repository, the merge rate fell by 1.2 percentage points (\textminus{}2.3 to \textminus{}0.02), which was statistically significant. With other pooling methods, the calendar comparison, or other outcome definitions, the change was between \textminus{}1.4 and 0.0 percentage points (Table S5). Developers in the calendar comparison who did not switch also had slightly lower merge rates after the releases. In the six upgrades with SWE-Bench Pro scores for both models, the new model scored 5 to 18 points higher on SWE-Bench Pro, but none of the changes in merge rate were statistically significant, and five of the six point estimates were negative (Figure 3).

\begin{figure*}[t]\centering\includegraphics[width=\linewidth]{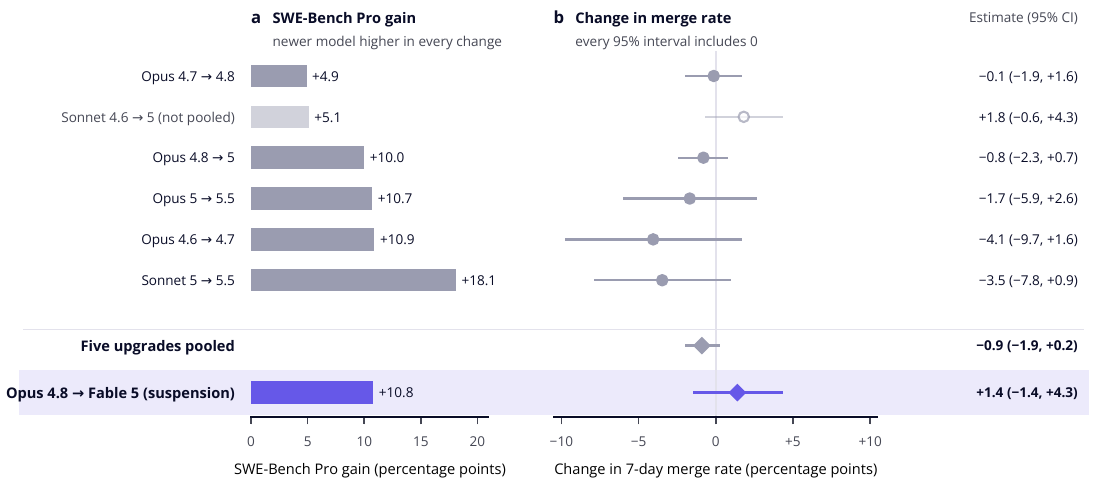}

\caption*{\textbf{Figure 3.} Benchmark gains of 5 to 18 points came with no rise in merging. a, SWE-Bench Pro gain of the newer model over the older one, from Anthropic's reported scores. b, Change in the same developers' 7-day merge rate, with 95\% intervals. For each upgrade, we used the PRs that developers who switched opened in the 7 days before and after the release, in linear probability models with developer fixed effects. The five upgrades in the main analysis are pooled by inverse-variance weighting (Appendix A). The light bar and open circle show Sonnet 4.6 to 5, which had too few developers who switched to be pooled. The suspension row is the reversal contrast, with developer and calendar-day fixed effects. Intervals are clustered by developer. The two February upgrades are not shown, because SWE-Bench Pro scores are not available for both models. All eight upgrades are in Table S3.}\end{figure*}

We also looked at whether developers got more PRs merged per week after switching. For each of the six upgrades, we collected all PRs with a Claude co-author line that developers who switched and developers who did not switch opened in the week before and the week after (Table S2). Both groups opened fewer PRs in the week after the release. Compared with developers who did not switch, developers who switched got 2.4\% more PRs merged within 7 days per week (\textminus{}4.9\% to +10.2\%), which was not statistically significant. They also opened 2.7\% more PRs (\textminus{}6.7\% to +13.0\%).

The treated group also did not slow down during the suspension. Counting all PRs they opened, with or without a Claude line, each developer opened 4.8 PRs per day before the release, 5.0 during the suspension, and 5.2 after access came back (Appendix F). The upgrade comparison above is a better test of weekly merged PRs, because it selects both groups in the same way.

We also looked at whether PRs from the new models failed automated checks less often. Among the PRs that ran checks, the share with a failed check changed by +0.3 percentage points after switching (95\% CI \textminus{}1.4 to +2.0; Table 2), which was not statistically significant, i.e., switching to a new model reduced check failures by at most 1.4 percentage points (Appendix A).

\begin{table*}[t]\centering

\caption*{\textbf{Table 2.} Merge outcomes and check failures did not change with access to Fable 5 or after switching to a newer model. Estimates with 95\% confidence intervals. The suspension analysis gives the reversal contrast (the two phases with Fable 5 minus the suspension) from the linear probability model with developer and calendar-day fixed effects (297 treated and 8,908 control developers). The upgrade analysis gives the change after switching, with developer fixed effects, pooled over the five upgrades by inverse-variance weighting (1,208 developers). Standard errors are clustered by developer. Percent changes come from log(1 + x). \textsuperscript{a}Compared with developers who did not switch, pooled over six upgrades (Appendix A). \textsuperscript{b}The suspension changed which PRs name a model, so weekly counts of named PRs are not comparable across phases (Appendix F). \textsuperscript{c}16 reverts before and 2 after the switch (Appendix F). \textsuperscript{d}Among PRs that ran automated checks, pooled over the five upgrades. Not estimated for the suspension analysis, which has no comparison group for this outcome (Appendix A). Full estimates are in Table S5.}

\small

\begin{tabular}{l cc}\toprule

Outcome & Suspension analysis & Upgrade analysis\\

\midrule

\multicolumn{3}{l}{\textit{Main outcomes}}\\

Merged within 7 days (percentage points) & +1.4 (\textminus{}1.4, +4.3) & \textminus{}0.9 (\textminus{}1.9, +0.2)\\

Weekly merged PRs (\%) & not estimated\textsuperscript{b} & +2.4 (\textminus{}4.9, +10.2)\textsuperscript{a}\\

\multicolumn{3}{l}{\textit{Secondary outcomes}}\\

Merged within 1 hour (percentage points) & 0.0 (\textminus{}5.8, +5.8) & +0.8 (\textminus{}2.1, +3.8)\textsuperscript{a}\\

Closed without merge (percentage points) & \textminus{}0.5 (\textminus{}3.0, +2.1) & +0.5 (\textminus{}1.3, +2.3)\textsuperscript{a}\\

Comments per PR & not estimated & +0.03 (\textminus{}0.06, +0.12)\textsuperscript{a}\\

Reverted within 30 days (percentage points) & 0.0 (\textminus{}0.4, +0.4) & too few reverts\textsuperscript{c}\\

Failed automated check (percentage points) & -- & +0.3 (\textminus{}1.4, +2.0)\textsuperscript{d}\\

Failed test or build check (percentage points) & -- & +0.2 (\textminus{}1.6, +1.9)\textsuperscript{d}\\

\multicolumn{3}{l}{\textit{Checks on the merge rate}}\\

Placebo: before the release vs suspension & \textminus{}2.6 (\textminus{}6.2, +1.1) & --\\

Within developer and repository & +1.4 (\textminus{}1.4, +4.3) & \textminus{}1.2 (\textminus{}2.3, \textminus{}0.02)\\

Refit on all PRs, with or without a model name & +0.5 (\textminus{}2.3, +3.3) & --\\

Time to merge among merged PRs (\%) & +8.4 (\textminus{}9.0, +29.2) & +2.4 (\textminus{}2.1, +7.2)\\

\bottomrule\end{tabular}

\end{table*}

\subsection{Almost all merges were done by the authors themselves}

We found that of the 2,019 sampled PRs in the suspension analysis, 95\% were merged by the author, and 90\% had no approval from any other account (Appendix C). Authors merged their own PRs whether or not they had a public relationship with the repository (97\% of merged PRs by inside authors and 94\% by outside authors). In every phase of the suspension analysis and in every week around each upgrade, the median time from opening a PR to merging it was under half an hour, regardless of model. Within the same developer, time to merge changed by +8.4\% (\textminus{}9.0\% to +29.2\%) when Fable 5 was available and by 2.4\% (\textminus{}2.1\% to +7.2\%) after the upgrades, which were not statistically significant (Table S5).

Almost all PRs were merged even before the upgrades. Across the six upgrades, in the week before each upgrade, developers who switched got 94.5\% of their PRs merged within 7 days and 77.1\% within one hour, and 83.8\% of the 1,262 developers who switched had every one of their PRs merged. Because almost all PRs were already merged, a new model could raise the merge rate only for developers who had some unmerged PRs. Among developers with at least one unmerged PR in the week before the upgrade, the merge rate of those who switched changed by 1.3 percentage points lower compared with those who did not switch (\textminus{}6.9 to +4.3), which was not statistically significant.

\section{Discussion}

Model companies use benchmarks to advertise their models. Users pick models based on the benchmarks too~\cite{r34}, but no one has shown that these benchmarks actually reflect real work. Using the 19-day suspension of Fable 5 as a natural experiment, we found that Fable 5, which led by about 11 points on SWE-Bench Pro, did not make the PRs of the same developers more likely to be merged. We also found the same result in five other upgrades where developers chose which models to use.

Our findings are in the same direction as a previous study that compared GitHub developers in Italy with those in other European countries after Italy banned ChatGPT, where the overall output of experienced developers did not change~\cite{r40}, although they studied the removal of ChatGPT entirely, not a switch between models. Our findings differ from two studies of newer models: researchers found that contributions to open-source Python packages rose after GitHub Copilot's June 2022 model upgrade~\cite{r30}, and that each year of model progress cut task time by about 8\%~\cite{r39}. Nonetheless, these studies did not follow the same developers on work they chose themselves.

We hypothesize that the main reason Fable 5's gains on benchmarks did not translate to real-work productivity is that the tasks developers give agents are already within what the older model can do. A survey of 132 engineers and researchers at Anthropic found that they mostly give Claude tasks that are easy to check and cheap to get wrong, and more than half of them could fully hand off only 0 to 20\% of their work~\cite{r32}. Experienced developers also give agents clearly described, simple tasks~\cite{r31}. So the hard problems the new model solves on SWE-Bench Pro are not the work developers usually hand off. This also explains why almost all PRs were already merged before the upgrades, and why the merge rate and weekly merged PRs did not change. Developers also did not give the new model harder tasks, since PR description length, the number of files mentioned and the share of new features did not change after the upgrades (Appendix F).

Besides its 10.8-point lead on SWE-Bench Pro, Fable 5 led Opus 4.8 by 6.4 points on SWE-bench Verified~\cite{r6,r20,r44}. We also suspect Fable 5's gains on coding benchmarks did not show up in merged work because the benchmarks failed to represent real work. SWE-Bench Pro, for example, counts a task as solved if the code change passes the tests from the original fix~\cite{r43}. Apart from these tests having poor quality~\cite{r25}, they also only test a small portion of an engineer's work; they do not test other aspects required by the work, e.g., working out what a vague request needs or revising the code after review~\cite{r24,r33}. Model releases also do not report confidence intervals with benchmark gains, so it is unclear if they are larger than chance~\cite{r20,r28}. On Terminal-Bench 2.1, Fable 5 led Opus 4.8 by only 1.6 points, but this is not a sign of Terminal-Bench's tasks better matching real work; it is largely because one in five Fable 5 runs fell back to Opus 4.8 after a safety refusal~\cite{r20}. Terminal-Bench also does not aim to represent everyday work. It selects hard tasks inspired by real workflows by design~\cite{r42}. As a result, a higher score on Terminal-Bench also doesn't mean a model would help more in real work.

Our findings add evidence from real work to what prior work has warned about: a benchmark score means little to people choosing a model without evidence that it measures the work it is used to describe~\cite{r26,r27}. If regulators and buyers judge how much stronger a model is by its benchmark score~\cite{r34,r36}, the difference they rely on will likely not show up in user productivity.

Our outcome, merge rate, despite being used by many studies to represent coding productivity~\cite{r3,r4,r9,r21,r22}, is not a perfect proxy for productivity. We found that almost all merges were done by the authors themselves, which agrees with findings that agent PRs are often merged without review~\cite{r10}, so a merge only shows that the developer considered the work done, not code quality. If differences between models show up mainly when someone else reviews the work, our estimates are too small. However, a merge is still the best public signal we have of whether a developer decided the code was good enough to keep. To measure time saved, we would need a randomized trial, but such trials have other problems. For example, developers know they are being studied and may work differently, often do not use the tool they are assigned~\cite{r21}, and usually work on tasks given by the study instead of their own (Appendix A). Comparing merge rates across developers can also mislead. When we pooled all PRs that named Fable 5 or Opus 4.8 on seven days from July to August, PRs naming Opus 4.8 were merged 2.6 percentage points more often, but within the same developer and repository, the difference was only 0.9 points and not statistically significant (Appendix E).

We also found that almost all PRs were merged even before the upgrades, so the merge rate had little room to rise. But merging within one hour, which started from about 75\%, did not change either (Table 2), nor did the merge rate among developers who had unmerged PRs (Section 4.3) or weekly merged PRs. The suspension lasted only 19 days, so the suspension analysis alone has wide intervals, but the results of the five upgrades agree with it and have much narrower intervals. In addition, we could only use the PRs whose descriptions name the model, which are about 1\% of merged Claude Code PRs, and this share fell during the suspension. However, these PRs were merged about as often as other Claude Code PRs, and when we refit the model on all PRs of the same developers, we got results in the same direction (Appendix B). The upgrade analysis is observational. Developers who switch models on their own may already be better at using new tools, which would bias the estimate upward, but even so, the merge rate did not rise. One possible reason the merge rate fell after the Opus 4.7 release is a quality problem in Claude Code from April 16 to 20, 2026, which affected both Opus 4.6 and 4.7 in the week right after the release~\cite{r23}. Other limitations are in Appendix H.

\section{Conclusion}

We used the 19-day suspension of Fable 5 and five model upgrades in Claude Code to test if having access to models scoring higher on coding benchmarks leads to more work done. Results show that having access to Fable 5 increased the 7-day merge rate by only 1.4 percentage points, and switching to a new model changed it by \textminus{}0.9 points (\textminus{}1.9 to +0.2), neither of which was statistically significant. The share of pull requests with failed automated checks also did not change after the upgrades (+0.3 points, \textminus{}1.4 to +2.0). This also shows that benchmark gains of 5 to 18 points did not show up in the work developers actually kept.

\section{Data and code}

This paper does not name any developer. We release the dataset we collected, with GitHub logins, repository names and PR links replaced by salted hash codes, and the code that reproduces every number and figure in this paper.\footnote{\url{https://github.com/ningkko/benchmarks-vs-merges}}

\clearpage\onecolumn\appendix\section*{Appendix}

\FloatBarrier\section{Design and analysis details}

Fable 5 was released on June 9, 2026~\cite{r12}, and Anthropic received the export control directive at 5:21 p.m. ET on June 12~\cite{r5}. The Fable 5 and Mythos 5 system card reports 80.0\% on SWE-Bench Pro for Fable 5 (80.3\% for Mythos 5, the same model), and the Opus 4.8 system card reports 69.2\% for Opus 4.8~\cite{r6,r20} (10.8 points difference compared to Fable 5). With 731 public tasks, the 95\% confidence interval of the 10.8-point lead is about 6.4 to 15.2 points if the two models' results are treated as independent.

The suspension analysis assumes that without the suspension, the day-to-day changes in the treated group's merge rate would have been the same as the control group's. To break this assumption, a confounder would have to affect only Fable 5 users, start exactly on June 12 and end on July 1. The treatment is also more than a benchmark gap. Losing Fable 5 also meant switching models, in which case speed and usage limits may have changed too, so we estimate the overall effect of losing and regaining Fable 5.

This overall effect is the comparison that benchmark scores are used to support. Users choose a model by its score, and they get the whole model with its speed and usage limits, not only the part the benchmark measures. The upgrades, which involved no interruption, gave the same result as the suspension.

During the suspension, many Fable 5 users moved to Opus 4.8, which could have made Opus 4.8 slower or changed its limits. The control group used Opus 4.8 throughout, so such a change affected both groups and is removed by the calendar-day fixed effects. It would bias the reversal contrast only if it affected Opus 4.8 users who came from Fable 5 differently from other Opus 4.8 users.

We left out the weekend after the suspension (June 12, 21:00 to June 15) and the day of the restoration (July 1), because PRs opened right after a switch may contain code written before it (phase times in Table S2). P1 lasts only three days and P3 two weeks, so the two are not symmetric around the suspension. If the treated group's merge rate has a steady time trend, the reversal contrast does not cancel it, so we also refit with a weekly linear trend for the treated group (Table S5).

The two weeks before the release also support the assumption. The treated group's merge rate in these weeks differed from the reference week by \textminus{}1.5 (\textminus{}9.0 to +6.0) and \textminus{}2.5 (\textminus{}8.8 to +3.8) percentage points, close to each other and to 0 (Figure 2). The placebo compares two phases without Fable 5, before the release and the suspension, so it shows whether the two groups moved together when nothing changed for the treated group. It was \textminus{}2.6 (\textminus{}6.2 to +1.1). With the weekly linear trend, the reversal contrast was +1.6 (\textminus{}1.0 to +4.3; Table S5).

Table S3 lists the eight upgrades and their release dates [6--8, 13--17]. PRs still given to the old model after the release are not in the sample. An upgrade enters the pooled estimate, with inverse-variance weights, only if at least 10 developers who switched opened outside-author PRs in both weeks, because the interaction in Appendix C needs these PRs. Six upgrades meet this rule, but the February Sonnet upgrade and Sonnet 4.6 to 5 do not. We had already looked at the data of the February Opus upgrade in an earlier exploratory analysis, so the merge rate pools only the other five, and weekly merged PRs and the secondary outcomes in Table 2 pool all six.

The rule counts developers, not outcomes, so it cannot favor a result, and other choices give the same answer: \textminus{}0.8 (\textminus{}1.9 to +0.2) for six upgrades, \textminus{}0.4 (\textminus{}1.4 to +0.6) for all eight, and \textminus{}1.4 (\textminus{}2.9 to +0.1) in one stacked model (Table S5). Sonnet 4.6 to 5, which did not meet the rule, had a positive estimate (+1.8, \textminus{}0.6 to +4.3), but its week after the release includes the days when Fable 5 came back.

The change of the same developers around an upgrade also includes changes shared by everyone at the time, and a new model may raise activity simply because it is new. Subtracting the change of developers who did not switch gave \textminus{}1.1 (\textminus{}3.1 to +0.9), and 0.0 (\textminus{}2.5 to +2.5) with equal weight per developer, so even the widest interval rules out gains above about 2.5 points. For three upgrades this comparison group has only 17 to 27 developers (Table S4), and for Sonnet 5 to 5.5 the comparison model Opus 5.5 had been released only six days earlier.

Our collection follows AIDev v5~\cite{r1}, which holds 7,685,281 PRs from six agents up to March 31, 2026, 1,567,100 of them from Claude Code. Our queries are the same as AIDev's, with the model line added as a phrase filter, so the windows before March 31 reproduce data already in AIDev v5. After the search, we matched the model line exactly with a regular expression to remove loose matches, e.g., ``Generated-by: Claude Code opus-4.5'' in an Apache Spark template.

We only study Claude Code because the natural experiment only exists for Fable 5, a Claude model. The suspension removed Fable 5 from all customers for 19 days, set from outside regardless of what users wanted, and we know of no such change for other models. Claude Code also writes the model name into commits by default~\cite{r19}, which lets us see which model each PR came from.

No single outcome measures coding productivity. Studies of developer productivity have used lines of code, then counts of pull requests and merges~\cite{r21,r22}, and benchmarks use test pass rates, which do not come from real work at all. A merge is the closest public record of work that a developer accepted into the project, which is why many studies use it~\cite{r3,r4,r9,r21,r22}. Because almost all PRs were already merged, we also looked at outcomes with room to move. Merging within one hour started from about 75\% and did not change, either with access to Fable 5 (0.0, \textminus{}5.8 to +5.8) or after the upgrades (0.0, \textminus{}1.9 to +1.8; Table S5). Among developers with at least one unmerged PR in the week before the upgrade, the merge rate of those who switched changed by \textminus{}1.3 percentage points compared with those who did not switch (\textminus{}6.9 to +4.3; Section 4.3). Weekly merged PRs, which have no upper limit, changed by +2.4\% (\textminus{}4.9\% to +10.2\%), and the share closed without merge, comments and reverts did not change either (Table 2). We cannot see fixes after a merge that do not go through a revert.

Merges do not show whether the code worked, so we also looked at failed automated checks (Section 3.4). We collected the checks in November 2026, when 1.6\% of the PRs were no longer on GitHub. Of the rest, 85\% ran checks, and 19\% of these had a failed check. Comparing the same developers' PRs before and after each upgrade, the share with a failed check changed by +0.3 percentage points (95\% CI \textminus{}1.4 to +2.0), and by +0.2 (\textminus{}1.6 to +1.9) when we only counted test and build checks, which 58\% of PRs had (Table 2). Only one upgrade changed significantly, Opus 4.7 to 4.8, with more failures after the switch (+5.4, +1.1 to +9.7). Without fixing the developer, the failure rate seemed to drop after some upgrades (e.g., from 22.8\% to 11.0\% after Opus 5), but this came from which developers opened PRs after the upgrade, not from the model. During the suspension, test and build failures were 6.4 percentage points higher when Fable 5 was available (+0.1 to +12.8). However, this estimate used only the 2,019 sampled PRs of the treated group with no comparison group, and the failure rate changed a lot between phases (8\% to 40\%), so we only report it as a supplement.

Harder tasks and time saved could be measured only in a randomized trial, but trials have their own problems. Developers do not always use the tool they are assigned. In one large field experiment, only 44\% of developers given an AI coding tool at one company adopted it in the first 29 weeks, so the trial had to estimate the effect of assignment and recover the effect of use with further assumptions~\cite{r21}. Trials also often set the tasks, so they do not show how developers use agents on their own work, and trials that measure time on real tasks are small, e.g., 16 developers~\cite{r11}. Our design uses developers' own work at scale, with the change in model set from outside.

We use a linear probability model for three reasons. First, the coefficients are directly differences in merge rate in percentage points, the same unit as a leaderboard. Second, a logistic model with developer fixed effects drops every developer whose PRs were all merged (or all unmerged), and such developers are the majority. Third, the reversal contrast is an average of several coefficients, which is simple to compute in a linear model. When the merge rate is close to 100\%, some predicted values of a linear probability model can be above 1. But we estimate average differences, and standard errors clustered by developer are robust to the heteroskedasticity of this model.

Besides equal weight per PR, Table S5 reports other weightings: (1) at most 20 or 50 PRs per developer, (2) equal weight per developer, and (3) computing each developer's mean first and then pooling (2,000 bootstrap samples of developers). We estimate the developer-level intraclass correlation with one-way analysis of variance (Appendix E).

The main analysis gives each PR equal weight, because the merge rate is defined over PRs and our question is about the work done, so developers who open many PRs count more. Giving each developer equal weight answers a different question, the effect for a typical developer, and makes the estimate much less precise because most developers open few PRs: +3.7 (\textminus{}5.5 to +12.9) for the suspension. Capping each developer at 20 or 50 PRs gave +3.6 (\textminus{}1.1 to +8.3) and +3.4 (\textminus{}0.7 to +7.5; Table S5). For a typical developer, the gain may be a few points larger than for the average PR, but no weighting gave a point estimate above 4 points, and only one of them was statistically significant.

\FloatBarrier\section{Which PRs name the model}

The model line names the model in use when the PR was created, but the code in the PR may come from an earlier session with another model. To check this, we randomly sampled 200 merged PRs that name a model from the upgrade analysis (100 before and 100 after the upgrades), and searched GitHub for commits by the same author in the same repository. We could only match commits by author, repository and date, because the API that lists a PR's own commits is not open to bulk collection. We sampled only merged PRs, because commit search only indexes the default branch.

The first search window ran from one week before the PR was created to one day after it was merged. We found commits for 195 PRs, and for 152 of them, the commits named a model. For PRs created before the upgrade, 95\% matched the model in the commits, but for PRs created after, only 80\% did, and most mismatches named the older model. This is because for PRs after the upgrade, going back one week often reaches the time before the new model's release, when only the older model was available. So we moved the start to one day before creation and searched again. This time, the commits named a model for 138 PRs, and the match rate was 96\% before the upgrade and 99\% after. Among these 138 PRs, more commits after the upgrade also named another model (43\% vs 31\%), which means some work carried over from sessions with the older model.

We also checked how many Claude Code PRs name the model (Table S2). We sampled commits that name a Claude model on 8 days, from four random one-hour windows a day, and used the GitHub REST API to find the PR of each commit. Among 1,219 merged PRs randomly sampled from 8 days whose commits named a Claude model, only 1.0\% (95\% CI 0.6\% to 1.7\%) also named the model in the description. Another 76.7\% had only Claude Code's default line, and the rest had no Claude line. The share naming the model was between 0\% and 1.8\% in every suspension phase and in the upgrade days, and it stayed at 1.0\% without bot accounts. Because commit search only indexes the default branch, all of these PRs were merged, so this check cannot compare merge rates. So we compared merge rates with GitHub search counts. In each week, the merge rate of PRs naming a model differed from that of PRs with only Claude Code's default line by \textminus{}4.0 to +6.1 percentage points, and by \textminus{}1.5 to +1.4 in the suspension weeks (merged by the collection date, bot accounts excluded).

The PRs in this record do not look unusual, since their merge rate was close to that of PRs with only Claude Code's default line, and we know of no other public record that links PRs to a model. The treated developers adopted Fable 5 within its first three days, but before the release they were similar to the control group on all characteristics in Table 1 (standardized differences within 0.2), and the upgrade analysis, which uses a broader group of 1,208 developers, gave the same result. The results may not hold for workflows that do not carry commit messages into the PR description.

The share of PRs that name a model could also change with the model. Among the treated developers, the share of PRs with a Claude line fell during the suspension and only partly came back (Appendix F). If developers moved tasks that were easier or harder than usual to other agents during the suspension, the difference between phases would mix in a change of tasks. Any study of real work faces this, including a trial, because developers can move tasks to other tools during the study.

We checked this in several ways. When we refit the suspension model on every PR of the same developers, with or without a model name, the reversal contrast was +0.5 percentage points (\textminus{}2.3 to +3.3), and +1.0 (\textminus{}1.1 to +3.3) when we computed each developer's mean first (Appendix F). Fable 5 was only 14\% of P1 and 2\% of P3 in this sample, so this estimate is pulled toward 0, and we use it to check the direction. The kind of work stayed similar across phases, with standardized differences within 0.2 for the share of titles starting with ``feat'' and for description length (Table S6, Panel B). Only 52 developers opened PRs naming a model in both P2 and P3, but computing developer means separately for the suspension (84 developers) and the restoration (52 developers) gave +3.5 (\textminus{}2.5 to +9.7) and \textminus{}0.1 (\textminus{}6.9 to +7.4; Table S5), so no single group of developers drives the result. Finally, the upgrades did not change which PRs name a model (\textminus{}2.2 percentage points, \textminus{}4.9 to +0.4; Appendix F), and the five upgrades gave the same result as the suspension with much narrower intervals.

\FloatBarrier\section{Author association}

GitHub labels each PR with an author association. We call owners, members and collaborators inside authors, and all other authors outside authors, who have no public relationship with the repository. We expected PRs from outside authors to be merged by someone else, so we also estimated the interaction between the model switch and the author association, defined as the change for PRs by outside authors minus the change for PRs by inside authors. But PRs from outside authors were also mostly merged by the authors themselves (Section 4.3). The public API does not show private organization membership or collaborator permissions, which means many people labeled as outside authors in fact have merge permission.

In the 161 developers who opened PRs as outside authors in both weeks around an upgrade, the interaction was +0.2 percentage points (95\% CI \textminus{}3.5 to +3.8) pooled over five upgrades and +1.1 (\textminus{}2.4 to +4.6) pooled over six. Limiting inside authors to repository owners gave +0.2 (\textminus{}3.6 to +4.0) and +1.3 (\textminus{}2.4 to +4.9). Because outside authors mostly merged their own PRs, this interaction does not tell whether a new model helps more when someone else reviews the work.

We cannot tell whether a person or a script clicked the merge button. In both cases the author's account accepted the change into the project, which is what the merge rate records. Bot accounts are excluded (Section 3.2), and any automation a developer uses is removed by the developer fixed effect as long as it did not change when the model changed. The median time from opening to merging stayed under half an hour in every phase (Section 4.3).

\FloatBarrier\section{Estimates by actual use of Fable 5}

The reversal contrast in Section 4.1 estimates the effect of having access to Fable 5. But when it was available, developers did not use Fable 5 for every PR. In P1, 82\% of PRs named Fable 5, and in P3 only 27\% did. Compared with the suspension (0.9\%), the two phases are 54 percentage points higher on average. Dividing the reversal contrast by 0.54 gives +2.7 percentage points per PR that used Fable 5 (\textminus{}2.6 to +8.0). This scaling assumes, as in a standard instrumental-variable analysis, that access to Fable 5 changes the merge rate only through the PRs that use it. Using P1 alone gives +0.8 (\textminus{}4.6 to +6.2), and using P3 alone gives +8.5 (\textminus{}6.3 to +23.4).

We also directly compared the treated group's PRs that named Fable 5 with their other PRs (mostly naming Opus 4.8), and we call this the as-used comparison. Because developers chose which tasks to give Fable 5, this comparison is observational. Within the same developer, the difference was +2.7 percentage points (\textminus{}0.7 to +6.1), and within the same developer and repository, it was +3.0 (\textminus{}0.7 to +6.7). The share closed without merge changed by about the same amount in the other direction (\textminus{}2.0, \textminus{}4.9 to +0.9).

P3 has a special case. Another 33.7\% of the treated group's PRs in P3 carry a Claude line without a model name, and 82\% of these come from one developer's custom footer. The reversal contrast counts these as treated PRs, while the as-used comparison counts them as not naming Fable 5. In Table S5, the settings with at most 20 PRs per developer and with equal weight per developer limit this developer's influence on the estimates.

The size fields are too noisy to tell if developers gave Fable 5 larger tasks. In 1,262 sampled PRs from 54 developers who used both models, PRs naming Fable 5 changed about 16\% more lines (\textminus{}9\% to +48\%) and 4.1 more files (\textminus{}6.3 to +14.6) within the same developer. Across all 1,590 sampled PRs with size fields, the median number of changed lines was 107 and 114 for the two groups. Developers also wrote descriptions in similar ways. Within the same developer, description length differed by \textminus{}95 characters (mean 1,645), the number of files mentioned in the description by \textminus{}0.2 (mean 2.1), and the share of titles starting with ``feat'' by \textminus{}3 percentage points (mean 26\%).

\FloatBarrier\section{Pooled merge rates}

To see how a leaderboard that ranks by merge rate would rank the two models, we collected every PR naming Fable 5 or Opus 4.8 on June 10 and 11 (the last two full days before the suspension) and on each Wednesday from July 8 to August 19 (10,232 PRs in total). We took only one day per week because of the rate limits of the search API.

One day alone gives an unstable ranking. Fable 5 minus Opus 4.8 was +2.9 percentage points on June 10 (\textminus{}2.9 to +9.0) and \textminus{}4.2 on July 8 (\textminus{}7.4 to \textminus{}0.9).

Pooling the seven Wednesdays, we compared the two models in four ways (Figure S1). The first way works like a leaderboard and pools all PRs without adjustment. In this comparison, PRs naming Opus 4.8 were merged more often (91.7\% vs 89.2\%, a difference of \textminus{}2.6 percentage points, \textminus{}4.6 to \textminus{}0.5). The second way adds calendar-day fixed effects to the pooled comparison, and the difference was \textminus{}2.3 (\textminus{}4.2 to \textminus{}0.3). The third way compares only within the 226 developers who used both models, and the difference became smaller (\textminus{}2.1, \textminus{}4.9 to +0.6). The fourth way compares within the same developer and repository, and the difference became even smaller (\textminus{}0.9, \textminus{}3.5 to +1.7). The more the comparison is limited to the same person and the same repository, the smaller the lead of Opus 4.8.

\begin{figure}[htbp]\centering\includegraphics[width=\linewidth]{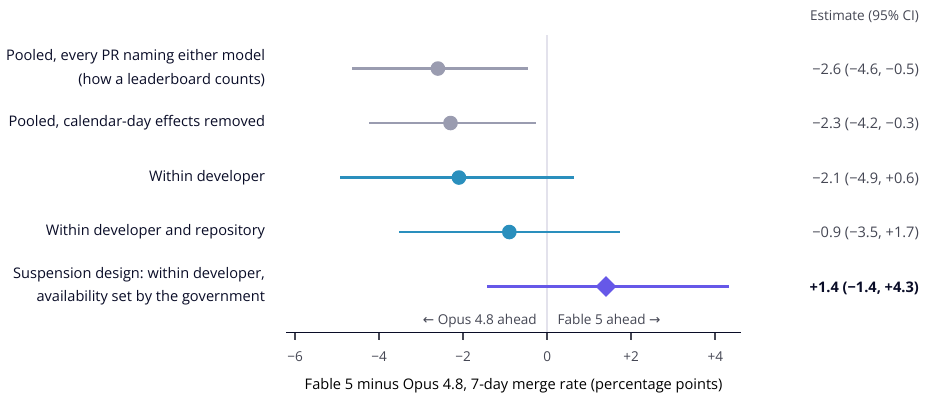}

\caption*{\textbf{Figure S1.} How the models are compared decides which one looks better. Fable 5 minus Opus 4.8 in 7-day merge rate, with 95\% intervals. The first four rows are the four comparisons in Appendix E (8,686 PRs on seven Wednesdays, July 8 to August 19, 2026), and the last row is the suspension design (Section 4.1). Pooled comparisons mix each model with the different developers who used it.}\end{figure}

This is because much of the variation in whether a PR is merged comes from differences between developers, including their repositories and review habits. The intraclass correlation of 7-day merging at the developer level is 0.42 in the suspension sample. So a comparison that pools PRs from different developers mostly compares the developers who chose each model.

\FloatBarrier\section{Changes in work, model naming and reverts}

\textit{Upgrade analysis.} Judging from PR descriptions, developers who switched did not give the new model different work after the upgrade. Compared with developers who did not switch, description length changed by \textminus{}75 characters (\textminus{}199 to +49), the number of files mentioned in the description by \textminus{}0.07 (\textminus{}0.33 to +0.19), and the share of titles starting with ``feat'' by \textminus{}0.9 percentage points (\textminus{}3.6 to +1.8). The upgrades also did not clearly change which PRs name the model. Compared with developers who did not switch, the share of descriptions that name the model and also carry Claude Code's default credit line changed by \textminus{}2.2 percentage points (\textminus{}4.9 to +0.4).

\textit{Workload during the suspension.} We collected every PR the treated developers opened from May 26 to July 16, with or without a Claude line, and did the same for 320 sampled control developers. Averaged over all 302 treated developers on the collection list, including those who opened none in the phase, each developer opened 4.8 PRs per day before the release, 5.0 during the suspension and 5.2 after access came back, so the suspension did not slow down their work. The number of PRs per control developer per day fell throughout the window (6.5, 4.1 and 3.4 in the three phases), which is related to how they were sampled. So a difference-in-differences on workload is not reliable (+8\% from the suspension to the restoration, \textminus{}2\% to +20\%; \textminus{}33\% from before the release to the suspension, \textminus{}41\% to \textminus{}25\%).

\textit{Fewer PRs named the model.} Of the 253 developers who opened PRs during the suspension, only 84 opened a PR with a Claude line. Among their PRs, the share with a Claude line fell from 16.5\% in the three release days to 6.7\% during the suspension (the same as before the Fable 5 release), and it was 8.1\% after access came back. In the three release days, developers did more work with the new model and named the model more often, so P1 is a short and unusual part of their work. The suspension estimate is based on only about 7\% of their PRs, and for the other PRs, we cannot see which agent wrote them or if an agent was used at all.

\textit{Refit on all PRs.} To avoid depending on which PRs named a model, we refit the suspension model on every PR of these developers (76,907 PRs from the 297 treated developers and 72,405 PRs from the 315 sampled control developers who opened PRs within the phase windows). The reversal contrast and placebo are in Table S5. When we computed each developer's mean first and then pooled, the phases when Fable 5 was available minus the suspension was +1.0 (\textminus{}1.1 to +3.3, 253 developers). In this sample, Fable 5 was only 14\% of P1 and 2\% of P3, so we cannot scale by use as in the first paragraph of Appendix D.

\textit{Reverts.} If one model's code is merged as often but breaks more often, we should see more reverts. For each repository in the suspension sample, we collected PRs with ``revert'' in the title and matched each one, by the original title or number, to a treated PR it reverted within 30 days of the merge. Of 5,878 merged treated PRs, 17 (0.3\%) were reverted within 30 days, and the share was 0.30\% both for PRs naming Fable 5 and for PRs naming Opus 4.8. In the five upgrades, 16 of the 7,471 PRs merged in the week before the switch were reverted within 30 days, and 2 of the 7,229 merged in the week after. The low count after the switch is partly because the last two upgrades had less than 30 days of follow-up (Table S2). These numbers are too small to interpret on their own, so we report estimates with ``merged within 7 days and not reverted within 30 days'' as the outcome in Table S5.

\textit{Type of work in the suspension analysis.} Many developers start PR titles with a label for the kind of change, and ``feat'' marks a new feature, as opposed to, e.g., ``fix'' for a bug fix or ``docs'' for documentation. We use the share of titles starting with ``feat'' to measure how much of the work is new features, and the length of the PR description as a rough measure of how much work the PR describes. Among all PRs the treated developers opened, with or without a model name, the share of titles starting with ``feat'' was 26.9\% in P1, 22.5\% in P2 and 20.6\% in P3, and the median description length was 1,273, 1,506 and 1,683 characters (all standardized differences within 0.2; Table S6, Panel B).

\FloatBarrier\section{Sensitivity analyses}

Table S5 lists the estimates of all checks (percentage points with 95\% CI unless noted). Models are as in Section 3.5, and the rows of the calendar comparison also add calendar-day fixed effects. Time to merge is the within-developer change in log(1 + hours) for merged PRs, converted to a percentage. In the suspension model, reverts were collected only for the treated group's repositories, so control PRs all count as not reverted. The two rows of developer means have no calendar adjustment, and their intervals come from 2,000 bootstrap samples. A blank cell means the check has no placebo. No reversal contrast in Panel A was above 4 percentage points, and no change in merging after the upgrades in Panel B was above 0.

Our question is whether the benchmark gains show up in merged work, so a null result is the finding, and what matters is how large a gain the data can rule out. The effects were small and not statistically significant. Even under the most favorable reading, the data rule out gains larger than about 4 percentage points with access to Fable 5 and larger than 0.2 points after an average upgrade, and the upper bound per PR that actually used Fable 5 is +8.0 (Appendix D). Benchmark points and merge-rate points come from different tasks with different baselines, so we do not compare the 11-point lead with these numbers directly.

\begin{table}[htbp]\centering

\caption*{\textbf{Table S5.} Estimates of all checks (percentage points with 95\% CI unless noted).}

\par\medskip\textit{Panel A: Suspension analysis}\par\smallskip

\small

\begin{tabular}{>{\raggedright\arraybackslash}p{9.2cm} cc}\toprule

Check & Reversal contrast & P0 placebo\\

\midrule

Main estimate (equal weight per PR, merged within 7 days) & +1.4 (\textminus{}1.4, +4.3) & \textminus{}2.6 (\textminus{}6.2, +1.1)\\

Weighting &  & \\

Equal weight per developer & +3.7 (\textminus{}5.5, +12.9) & \textminus{}2.0 (\textminus{}13.7, +9.8)\\

At most 20 PRs per developer & +3.6 (\textminus{}1.1, +8.3) & \\

At most 50 PRs per developer & +3.4 (\textminus{}0.7, +7.5) & \\

Developer means: suspension (n = 84) & +3.5 (\textminus{}2.5, +9.7) & \\

Developer means: restoration (n = 52) & \textminus{}0.1 (\textminus{}6.9, +7.4) & \\

Model &  & \\

Weekly linear trend for the treated group & +1.6 (\textminus{}1.0, +4.3) & \\

developer $\times$ repository fixed effects & +1.4 (\textminus{}1.4, +4.3) & \textminus{}2.6 (\textminus{}6.2, +1.1)\\

Sample: all PRs of the developers (with or without a model name) &  & \\

Equal weight per PR & +0.5 (\textminus{}2.3, +3.3) & \textminus{}1.6 (\textminus{}4.2, +1.1)\\

Control group weighted by entropy balancing & +0.4 (\textminus{}2.4, +3.3) & \\

Equal weight per developer & +3.6 (+0.1, +7.1) & \textminus{}0.2 (\textminus{}3.1, +2.8)\\

Other outcome definitions &  & \\

Merged within 1 hour & 0.0 (\textminus{}5.8, +5.8) & +7.8 (\textminus{}2.4, +18.0)\\

Merged within 1 day & +0.5 (\textminus{}4.1, +5.1) & \textminus{}2.9 (\textminus{}7.2, +1.5)\\

Merged within 28 days & +0.6 (\textminus{}2.1, +3.3) & \textminus{}3.2 (\textminus{}6.7, +0.3)\\

Merged within 7 days and not reverted within 30 days & +1.5 (\textminus{}1.6, +4.5) & \textminus{}2.5 (\textminus{}6.1, +1.2)\\

Closed without merge & \textminus{}0.5 (\textminus{}3.0, +2.1) & \\

Reverted within 30 days & 0.0 (\textminus{}0.4, +0.4) & \\

Time from creation to merge (\% change) & +8.4\% (\textminus{}9.0\%, +29.2\%) & \\

\bottomrule\end{tabular}

\par\medskip\textit{Panel B: Upgrade analysis}\par\smallskip

\small

\begin{tabular}{>{\raggedright\arraybackslash}p{11cm} c}\toprule

Check & Pooled change\\

\midrule

Main estimate (five upgrades, merged within 7 days) & \textminus{}0.9 (\textminus{}1.9, +0.2)\\

Pooling & \\

Six upgrades & \textminus{}0.8 (\textminus{}1.9, +0.2)\\

All eight upgrades & \textminus{}0.4 (\textminus{}1.4, +0.6)\\

Five upgrades in one model, SE clustered by developer across upgrades & \textminus{}1.4 (\textminus{}2.9, +0.1)\\

Calendar comparison & \\

Change for developers who did not switch & \textminus{}0.7 (\textminus{}1.7, +0.3)\\

Minus control & \textminus{}1.1 (\textminus{}3.1, +0.9)\\

Minus control, equal weight per developer & 0.0 (\textminus{}2.5, +2.5)\\

Subsample and interaction & \\

Developers with at least one unmerged PR in the week before (minus control) & \textminus{}1.3 (\textminus{}6.9, +4.3)\\

Interaction: outside minus inside authors (Appendix C) & +0.2 (\textminus{}3.5, +3.8)\\

Within repository & \\

developer $\times$ repository fixed effects & \textminus{}1.2 (\textminus{}2.3, \textminus{}0.02)\\

Sensitivity: adjusted for the author's own repository & \textminus{}0.9 (\textminus{}2.0, +0.2)\\

Sensitivity: adjusted for outside author & \textminus{}0.9 (\textminus{}2.0, +0.1)\\

Other outcome definitions & \\

Merged within 1 hour & 0.0 (\textminus{}1.9, +1.8)\\

Merged within 1 day & \textminus{}0.8 (\textminus{}2.1, +0.6)\\

Merged within 28 days & \textminus{}0.3 (\textminus{}1.3, +0.8)\\

Merged within 7 days and not reverted within 30 days & \textminus{}0.6 (\textminus{}1.7, +0.4)\\

Time from creation to merge (\% change) & +2.4\% (\textminus{}2.1\%, +7.2\%)\\

\bottomrule\end{tabular}

\end{table}

\FloatBarrier\section{Limitations in detail}

Code in a PR may come from an earlier session with another model (Appendix B). For the control group and the upgrade analysis, we removed PRs naming two models, and we also left out the days right after each switch. A mix of two models in one PR usually makes the difference smaller. But if wrong model lines differ between merged and unmerged PRs, the estimate could be biased in either direction.

Most fields are read from search results. The APIs that report PR size and the merging account are not open to bulk collection, so we read these two kinds of fields only for the 2,019 sampled PRs. As a result, the main analysis can only use whether the repository belongs to the author and the author association. A complete version should read these fields for every PR from the GitHub API.

The restoration phase overlaps with the Sonnet 5 release on June 30, the U.S. July 4 holiday, and the rule that subscription users can use at most half of their weekly limit on Fable 5 until July 7~\cite{r18}. So after July 1, developers may have kept Fable 5 for certain tasks. The as-used comparison (Appendix D) does not depend on the phases and gives a similar estimate (+2.7 percentage points), but because developers chose which model to use for each task, it is observational.

Single upgrades are noisy. To detect a 3-percentage-point change with 80\% power, we need about 650 to 1,500 developers who switched (based on developer-level standard deviations of 27 to 42 percentage points in our sample), and a single upgrade rarely has that many. Only after pooling the five upgrades does the interval for the overall change narrow to about 1 point on each side.

\FloatBarrier\section{Other tables}

Table S1 gives GitHub search totals for each model line by month, before our regular-expression check. Small counts before a model's release are loose phrase matches or descriptions edited later. Each model line rises in the month of its release and falls when the next model comes out. For example, PRs naming Opus 4.6 went from 13 in January to 4,639 in February, while PRs naming Opus 4.5 fell from 2,348 to 782. So the model line follows which model developers actually used.

\begin{table}[htbp]\centering

\caption*{\textbf{Table S1.} GitHub search totals for each model line by month, before the regular-expression check.}

\small

\begin{tabular}{l rrrr}\toprule

Month & Sonnet 4.5 & Opus 4.5 & Opus 4.6 & Sonnet 4.6\\

\midrule

Apr 2026 & 458 & 95 & 6,547 & 2,620\\

Mar 2026 & 305 & 185 & 9,724 & 2,548\\

Feb 2026 & 1,648 & 782 & 4,639 & 514\\

Jan 2026 & 2,330 & 2,348 & 13 & 4\\

Dec 2025 & 1,987 & 654 & 32 & 3\\

Nov 2025 & 52 & 5 & 31 & 0\\

\bottomrule\end{tabular}

\end{table}

Table S2 lists the collection and analysis settings.

{\footnotesize\begin{longtable}{>{\raggedright\arraybackslash}p{3.2cm} >{\raggedright\arraybackslash}p{12.2cm}}

\caption*{\textbf{Table S2.} Collection and analysis settings.}\\\toprule

Item & Setting\\

\midrule\endhead

Source & GitHub REST search API (search/issues), 100 results per page, at most 30 requests per minute\\

Collection date & November 2, 2026 (UTC) for the main collections, with later checks dated in their rows\\

Model-line query & is:pr ``Co-Authored-By: Claude \textless{}model\textgreater{}'' in:body created:\textless{}window\textgreater{}\\

Developer histories & is:pr author:\textless{}login\textgreater{} ``Co-Authored-By: Claude'' in:body created:2026-05-26..2026-07-16\\

Windows & any window above 1,000 matches split into shorter windows (down to 30 minutes), so no window was truncated\\

Model-line check & regular expression for ``Co-Authored-By: Claude'', a family name (Opus, Sonnet, Haiku, Fable or Mythos) and a version; bodies naming two models dropped\\

Suspension phases (UTC) & P0 May 26 to Jun 9 18:00; P1 Jun 9 18:00 to Jun 12 21:00; P2 Jun 15 to 30; P3 Jul 2 to 15\\

Treated & 297 developers with a PR naming Fable 5 in P1; 6,352 PRs in P0 to P3\\

Control & 8,908 developers with PRs naming Opus 4.8 on any day from May 26 to July 15 and no PR naming Fable 5 in the June or July--August samples; 34,013 PRs in P0 to P3\\

Pooled sample & every PR naming Fable 5 or Opus 4.8 on Jun 10 and 11 and each Wednesday Jul 8 to Aug 19, 2026; 4,798 Fable 5 and 5,434 Opus 4.8 PRs after exclusions\\

Exclusions & bot authors; bodies naming two models; June 12 21:00 to June 15 and July 1\\

Outcomes & merged within 7 days of opening; closed without merge; comments; author owns the repository; author association\\

Models & linear probability, developer and calendar-day fixed effects (alternating projections, tolerance 1e-10), standard errors clustered by developer (CR1)\\

Bootstrap & 2,000 resamples of developers, seed 11, for developer-level means\\

Upgrade events & release dates and models in Table S3; older model 7 days before and newer model 7 days after each release, release day left out; inclusion rule at least 10 developers who switched with outside-author PRs in both windows; main pool events 3, 4, 6, 7 and 8\\

Calendar check & developers who did not switch named another model in both windows and never the event's models: Sonnet 4.5 (event 1), Sonnet 4.6 (events 3 and 4), Opus 4.8 (event 5), Sonnet 5 (events 6 and 7), Opus 5.5 (event 8); developer and calendar-day fixed effects\\

Other checks & description length and files named are missing for event 1\\

Volume check & every PR with a Claude co-author line that developers who switched and developers who did not switch opened from 7 days before to 7 days after each release, release day left out; several authors per search query; change in the log of PRs opened and of PRs merged within 7 days (plus one), developers who switched minus developers who did not switch, Welch standard errors\\

Commit check & random sample (seed 11) of 200 merged PRs naming a model from developers who switched, 100 opened before and 100 after an upgrade; GitHub commit search for the same author and repository from 7 days before opening to 1 day after merging, and again from 1 day before opening\\

Merger and size & every outside-author PR of the treated developers (1,019) and 1,000 sampled PRs by inside authors (seed 11); merged\_by, additions, deletions and changed\_files read from the GitHub PR endpoint; zero-valued fields are omitted by the endpoint and recorded as 0\\

Reverts & PRs with ``revert'' in the title in each repository of the treated sample, opened May 26 to Aug 31, 2026, matched to a merged treated PR by its title in quotes or by \#number within 30 days of the merge; the same for each repository of the merged PRs of developers who switched in events 3, 4, 6, 7 and 8, opened Apr 9 to Nov 12, 2026, collected Nov 4, so the last after-switch merges of events 7 and 8 had 22 to 30 days of revert coverage\\

Reviews & every review on the 2,019 sampled PRs from the PR reviews endpoint; a reviewer is another account when its login differs from the author's, and bot-like when the login ends in bot or [bot]\\

Event study & treated $\times$ period indicators in the suspension model; periods May 26 to Jun 1, Jun 2 to Jun 9 18:00, Jun 9 to 12 (Fable 5 on), Jun 15 to 21, Jun 22 to 28 (reference), Jun 29 to 30, Jul 2 to 8, Jul 9 to 15\\

All PRs & the suspension model on every PR of the 297 treated developers and the 320 sampled control developers from the workload check, with or without a model name, same phases and excluded days; one weight per PR and equal weight per developer\\

Share of PRs naming the model & part 1, commits whose message names a Claude model, sampled on 8 days (Jun 3, Jun 10, Jun 22, Jul 8, Apr 20, Jul 28, Sep 24 and Oct 1, 2026) from four random one-hour windows a day; each commit's PR found with the GitHub REST API (repos/\{repo\}/commits/\{sha\}/pulls), run Nov 12, 2026; part 2, weekly GitHub search counts of PRs naming a model and PRs with only Claude Code's default line, merged by the collection date, bot accounts excluded\\

\bottomrule\end{longtable}}

Table S3 lists every upgrade event (estimates in percentage points with 95\% CI). Anthropic reported both models' scores for six events: Opus 4.6 53.4 and Opus 4.7 64.3~\cite{r13}, Opus 4.8 69.2~\cite{r6}, Sonnet 4.6 58.1 and Sonnet 5 63.2~\cite{r17}, Opus 5 79.2~\cite{r14}, Opus 5.5 89.9~\cite{r15} and Sonnet 5.5 81.3~\cite{r16}. We found no score reported by Anthropic for Opus 4.5 or Sonnet 4.5. The interaction for event 2 is not shown, because only two developers who switched opened outside-author PRs, which is too few to estimate it. All five upgrades in the main pool had a negative change for developers who switched, and all five intervals included 0.

\begin{table}[htbp]\centering

\caption*{\textbf{Table S3.} Upgrade events (percentage points, 95\% CI).}

\scriptsize

\begin{tabularx}{\textwidth}{c >{\raggedright\arraybackslash}p{2.3cm} *{5}{>{\centering\arraybackslash}X} >{\raggedright\arraybackslash}p{1.3cm}}\toprule

Event & Release and models & PRs, older / newer model & Developers who switched (with outside-author PRs) & PRs of developers who switched & Change, all PRs & Interaction & Pool\\

\midrule

1 & Feb 5, Opus 4.5 to 4.6 & 677 / 1,112 & 54 (10) & 443 & +1.7 (\textminus{}5.2, +8.5) & +11.8 (\textminus{}0.8, +24.5) & six only\\

2 & Feb 17, Sonnet 4.5 to 4.6 & 578 / 270 & 9 (2) & 104 & +5.1 (\textminus{}51.3, +61.4) & too few to estimate & none\\

3 & Apr 16, Opus 4.6 to 4.7 & 2,473 / 2,795 & 115 (26) & 1,516 & \textminus{}4.1 (\textminus{}9.7, +1.6) & \textminus{}6.8 (\textminus{}29.6, +16.0) & five and six\\

4 & May 28, Opus 4.7 to 4.8 & 4,869 / 3,513 & 113 (24) & 3,482 & \textminus{}0.1 (\textminus{}1.9, +1.6) & \textminus{}2.9 (\textminus{}8.1, +2.4) & five and six\\

5 & Jun 30, Sonnet 4.6 to 5 & 3,676 / 4,037 & 193 (8) & 1,440 & +1.8 (\textminus{}0.6, +4.3) & +6.7 (\textminus{}9.8, +23.2) & none\\

6 & Jul 24, Opus 4.8 to 5 & 7,425 / 6,489 & 404 (47) & 5,477 & \textminus{}0.8 (\textminus{}2.3, +0.7) & +3.1 (\textminus{}3.9, +10.1) & five and six\\

7 & Sep 22, Opus 5 to 5.5 & 6,691 / 8,720 & 458 (50) & 4,000 & \textminus{}1.7 (\textminus{}5.9, +2.6) & \textminus{}2.5 (\textminus{}18.9, +14.0) & five and six\\

8 & Sep 28, Sonnet 5 to 5.5 & 4,777 / 1,796 & 118 (14) & 967 & \textminus{}3.5 (\textminus{}7.8, +0.9) & +6.2 (\textminus{}2.8, +15.2) & five and six\\

\bottomrule\end{tabularx}

\end{table}

Table S4 gives the estimates of the calendar comparison for each upgrade (percentage points, 95\% CI). Event 2 had too few developers who switched to compare. Developers who did not switch also had lower merge rates after most releases, so part of the decline among developers who switched was shared by everyone at the time. After subtracting their change, no upgrade showed a clear increase, and all intervals included 0.

\begin{table}[htbp]\centering

\caption*{\textbf{Table S4.} Calendar comparison for each upgrade (percentage points, 95\% CI).}

\scriptsize

\begin{tabularx}{\textwidth}{c l *{4}{>{\centering\arraybackslash}X}}\toprule

Event & Comparison model & Developers who did not switch (PRs) & Their change & Change for developers who switched & Calendar-controlled change\\

\midrule

1 & Sonnet 4.5 & 17 (129) & \textminus{}14.0 (\textminus{}31.4, +3.4) & +1.7 (\textminus{}5.2, +8.5) & +15.5 (\textminus{}0.8, +31.7)\\

3 & Sonnet 4.6 & 27 (260) & \textminus{}8.5 (\textminus{}18.8, +1.9) & \textminus{}4.1 (\textminus{}9.7, +1.6) & +5.4 (\textminus{}4.5, +15.4)\\

4 & Sonnet 4.6 & 27 (286) & +0.3 (\textminus{}16.3, +16.8) & \textminus{}0.1 (\textminus{}1.9, +1.6) & \textminus{}1.0 (\textminus{}17.1, +15.1)\\

5 & Opus 4.8 & 418 (4,991) & \textminus{}0.3 (\textminus{}1.9, +1.3) & +1.8 (\textminus{}0.6, +4.3) & +1.9 (\textminus{}0.9, +4.7)\\

6 & Sonnet 5 & 256 (2,213) & \textminus{}0.1 (\textminus{}2.3, +2.0) & \textminus{}0.8 (\textminus{}2.3, +0.7) & \textminus{}0.5 (\textminus{}3.0, +2.0)\\

7 & Sonnet 5 & 210 (1,911) & +0.2 (\textminus{}1.9, +2.3) & \textminus{}1.7 (\textminus{}5.9, +2.6) & \textminus{}2.4 (\textminus{}7.5, +2.7)\\

8 & Opus 5.5 & 424 (3,899) & \textminus{}1.1 (\textminus{}2.3, +0.2) & \textminus{}3.5 (\textminus{}7.8, +0.9) & \textminus{}3.4 (\textminus{}7.9, +1.1)\\

\bottomrule\end{tabularx}

\end{table}

Within the treated group, the 7-day merge rate stayed between 90.5\% and 94.6\% across phases, while merging within one hour and comments per PR changed more between P1 and P2 (Panel B). After switching, developers who switched opened fewer PRs as outside authors (from 23.5\% to 13.3\%; Panel C), so Table S5 also reports estimates adjusted for the author's own repository and for outside authors.

\begin{table}[htbp]\centering

\caption*{\textbf{Table S6.} Full version of Table 1. Data are all PRs these developers opened between May 26 and July 15, 2026, whether or not they name a model (297 treated developers and 320 sampled control developers). Panel A compares the two groups before the Fable 5 release (P0), and Panel B follows the treated group across the four phases. The standardized difference is the difference in means divided by the pooled standard deviation, and PRs per developer, comments and description length are on the log(1 + x) scale. In Panel B, each phase is compared with the suspension (P2). The weighted control group uses entropy balancing to match the treated group on the P0 merge rate, the share in their own repository and PRs per developer. Panel C compares the PRs that developers who switched opened in the week before (old model) and the week after (new model) each of the five upgrades in the main analysis, and here the standardized difference is the week after minus the week before.}

\par\medskip\textit{Panel A: Treated and control groups before the Fable 5 release (P0)}\par\smallskip

\scriptsize

\begin{tabularx}{\textwidth}{>{\raggedright\arraybackslash}p{4.4cm} *{5}{>{\centering\arraybackslash}X}}\toprule

 & Treated & Control & Weighted control & Std. diff. & Std. diff. (weighted)\\

\midrule

Developers with PRs in P0 & 244 & 315 & 312.6 (effective) &  & \\

PRs in P0 & 21,377 & 30,630 & 30,630 &  & \\

PRs per developer, median (IQR) & 34(10--88) & 35(10--89) & 32(10--84) & \textminus{}0.06 & \textminus{}0.02\\

Merged within 7 days (\%) & 92.7 & 91.5 & 92.7 & +0.05 & 0.00\\

Merged within one hour (\%) & 73.9 & 71.5 & 72.8 & +0.05 & +0.03\\

Closed without merge (\%) & 6.0 & 6.8 & 5.6 & \textminus{}0.03 & +0.02\\

Comments per PR, mean & 0.98 & 0.82 & 0.79 & +0.12 & +0.13\\

Opened in the author's own repository (\%) & 65.9 & 67.3 & 65.9 & \textminus{}0.03 & 0.00\\

Author association: member(\%) & 5.1 & 5.3 & 5.9 & \textminus{}0.01 & \textminus{}0.03\\

Author association: collaborator(\%) & 10.2 & 10.6 & 11.1 & \textminus{}0.02 & \textminus{}0.03\\

Author association: contributor(\%) & 18.6 & 16.0 & 16.5 & +0.07 & +0.05\\

Author association: none(\%) & 0.3 & 0.7 & 0.6 & \textminus{}0.07 & \textminus{}0.05\\

Title starts with ``feat'' (\%) & 23.5 & 31.5 & 31.0 & \textminus{}0.18 & \textminus{}0.17\\

Description length, median (characters) & 1,393 & 1,259 & 1,277 & +0.15 & +0.13\\

\bottomrule\end{tabularx}

\par\medskip\textit{Panel B: The treated group across phases}\par\smallskip

\scriptsize

\begin{tabularx}{\textwidth}{>{\raggedright\arraybackslash}p{3.4cm} *{7}{>{\centering\arraybackslash}X}}\toprule

 & P0 & P1 & P2 & P3 & Std. diff., P0 vs P2 & P1 vs P2 & P3 vs P2\\

\midrule

Developers with PRs & 244 & 297 & 253 & 234 &  &  & \\

PRs & 21,377 & 9,202 & 24,399 & 21,929 &  &  & \\

Merged within 7 days (\%) & 92.7 & 94.6 & 91.8 & 90.5 & +0.04 & +0.11 & \textminus{}0.04\\

Merged within one hour (\%) & 73.9 & 76.9 & 64.7 & 60.9 & +0.20 & +0.27 & \textminus{}0.08\\

Closed without merge (\%) & 6.0 & 4.1 & 6.9 & 7.6 & \textminus{}0.04 & \textminus{}0.12 & +0.03\\

Comments per PR, mean & 0.98 & 0.75 & 1.48 & 1.54 & \textminus{}0.22 & \textminus{}0.37 & +0.10\\

Opened in the author's own repository (\%) & 65.9 & 66.0 & 59.6 & 62.5 & +0.13 & +0.13 & +0.06\\

Title starts with ``feat'' (\%) & 23.5 & 26.9 & 22.5 & 20.6 & +0.02 & +0.10 & \textminus{}0.05\\

Description length, median (characters) & 1,393 & 1,273 & 1,506 & 1,683 & \textminus{}0.02 & \textminus{}0.10 & +0.19\\

\bottomrule\end{tabularx}

\par\medskip\textit{Panel C: Developers who switched in the upgrade analysis, before and after}\par\smallskip

\scriptsize

\begin{tabular}{>{\raggedright\arraybackslash}p{5cm} ccc}\toprule

 & Week before & Week after & Std. diff.\\

\midrule

Developers with PRs & 1,208 & 1,208 & \\

PRs & 7,829 & 7,613 & \\

PRs per developer, median (IQR) & 2 (1--6) & 2 (1--5) & \textminus{}0.04\\

Merged within 7 days (\%) & 95.0 & 93.7 & \textminus{}0.06\\

Merged within one hour (\%) & 77.8 & 77.1 & \textminus{}0.02\\

Closed without merge (\%) & 3.9 & 3.8 & 0.00\\

Opened in the author's own repository (\%) & 64.3 & 74.5 & +0.22\\

Outside author (\%) & 23.5 & 13.3 & \textminus{}0.26\\

Title starts with ``feat'' (\%) & 18.3 & 23.4 & +0.12\\

Description length, median (characters) & 1,267 & 1,098 & \textminus{}0.14\\

\bottomrule\end{tabular}

\end{table}

\end{document}